# Chapter 16

# Commissioning and Operation

*M. Lamont\*, M. Pojer and J.M. Jowett*

CERN, Geneva, Switzerland

## 16 Commissioning and operation

### 16.1 Commissioning and operation

At the point of commissioning and subsequent operation of the HL-LHC, the LHC itself will have been operational for over 10 years and a wealth of knowledge and experience will have been built up. The key operational procedures and tools will have been well established. The understanding of beam dynamics will be profound and refined by relevant measurement and correction techniques. Key beam-related systems will have been thoroughly optimized and functionality sufficiently enhanced to deal with most challenges up to that point. Availability will have been optimized significantly across all systems. This collected experience will form the initial operational basis following the upgrade.

However, the HL-LHC will pose significant additional challenges. The target integrated luminosity will make considerable demands on machine availability and operational efficiency. The planned beam characteristics will push beam dynamics to new limits. In the following discussion the operational cycle is revisited in light of the key HL-LHC challenges. The expectations and issues relating to availability in the HL-LHC are then outlined.

The planned bunch intensity, $\beta^*$, and compensation of the geometrical reduction factor lead to a potential bunch luminosity well above the acceptable maximum in terms of pile-up for the experiments. Thus, as discussed in Chapter 2, an obligatory operational principle of the HL-LHC will be luminosity levelling. The aim is to reduce a potential peak or virtual luminosity to a more manageable levelled value by some luminosity reduction technique. As the number of particles in the beam falls with time, appropriate adjustments keep the luminosity at the levelled value. The options include the use of transverse offsets, crab cavities, variation of the crossing angle, and dynamic change of $\beta^*$. A combination (e.g. dynamic change of $\beta^*$ and crab cavities) will be deployed. The longitudinal pile-up density is another critical parameter for the experiments, which must be taken into account during the levelling process.

#### 16.1.1 The nominal cycle

The nominal operation cycle provides the framework underpinning luminosity production. Given the higher than ever stored beam energy of the HL-LHC era, the nominal cycle must be fully mastered for effective, safe operation. As of 2012, the nominal operational cycle is well established for 50 ns and bunch population exceeding nominal. A brief outline of the phases of the nominal cycle follows.

- Pre-cycle. Following extensive experience during Run 1, it is known that the magnetic state of the machine can be re-established before every fill by rigorous application of either a combined ramp-down/pre-cycle of the main circuits and a de-gaussing of the corrector circuits, or, following an access, by a full cycle of all circuits.

- Set-up. The machine settings are verified with the injection of a limited number of bunches that have a reduced population (pilot, one bunch) or nominal population (intermediate, 12 bunches).

---

\* Corresponding author: Mike.Lamont@cern.ch

- Injection. The two LHC rings are progressively filled with trains of bunches transferred from the SPS to the LHC. The injection process should be well optimized by the time of the upgrade. However, high bunch intensities will have to be anticipated and all necessary mechanisms to minimize particle loss and emittance blow-up, and maximize reproducibility, should be deployed.

- Ramp. Once the machine is filled with the required number of bunches, the beams are accelerated to top momentum (up to 7 TeV/c). The ramp should hold no surprises. The characteristics of snap-back and correction of associated parameter swings can be taken as given. The 10 A/s ramp rate for main dipole power converters will still hold. Transverse feedback and tune feedback, and their healthy cohabitation, should be anticipated. Orbit feedback will be mandatory. Controlled longitudinal blow-up will be required.

- Squeeze. The mechanics of the squeeze including feedback will have to have been fully mastered. Issues of beam instabilities should have been resolved and might include squeezing with colliding beams. Again, this should have been made operational before the HL-LHC era. The possibility of combining the part of the squeeze process in the ramp is being considered to reduce the time required for the squeeze process.

- Adjust. At the end of the squeeze the beams are brought into collision at the interaction points. There is some flexibility in sequencing the order in which the beams are collided in the different interaction points. The process is relatively fast and will be well established.

- Stable beams. Once all the above procedures are completed and no abnormal conditions are detected, the conditions are met for safely switching on the detectors and the start of data taking. In general, this phase is expected to be stable with sufficient Landau damping from beam–beam, and small movements in beam overlap, thanks to good orbit stability. Emittance growth and losses to the collimation regions must be anticipated.

The principal challenges during the cycle involve injection, and transmission through the cycle of high bunch and high beam current, while preserving emittance. Given the experience of LHC operations thus far, the following potential issues may be identified:

- electron cloud;
- UFOs;
- localized beam-induced heating of specific hardware or vacuum assembly non-conformities – this can result in outgassing and bad vacuum as a result;
- cryogenic heat load;
- excessive beam loss in the cleaning sections.

High bunch population implies the need for very good parameter control and the need to properly control beam loss through the cycle. Beam instabilities will have to be fully under control using a variety of measures. Low $\beta^*$ implies strong non-linearities from the inner triplets and the need for associated correction. On the positive side, the HL-LHC should be able to count on excellent magnetic reproducibility and good stability. Hardware systems such as power converters, RF, and transverse dampers have already demonstrated impressive performance.

16.1.2   Availability and operational efficiency

The terms relevant to the year-long operational view are defined below.

- Scheduled proton physics time (SPT) is the time scheduled in a given year for high luminosity proton physics. It does not include initial re-commissioning, special physics runs, ions, machine development (MD), and technical stops. It does include the intensity ramp-up following re-commissioning at the start of the year. Note that high luminosity running involves a number of challenges not present in other modes of operation, and different machine availability can be expected.

- Availability is the scheduled proton physics time minus the time assigned to faults and fault recovery. It is normally expressed as a percentage of the SPT. Edge effects (e.g. recovery from access, the pre-cycle) should be fully included in the assigned fault time.
- The turnaround time is defined as the time taken to go from stable beam mode back to stable beam mode in the absence of significant interruptions due to fault diagnosis and resolution.
- Physics efficiency is the fraction of the scheduled physics time spent in stable beams.

### 16.1.2.1 Standard year of operation

The longer-term operational planning consists of a series of long years of operation (see Table 16-1) interspersed with long shutdowns of the order of a year or more. The long shutdowns are foreseen for essential plant maintenance, experiment upgrades, and so forth. It is estimated that the required length of a standard long shutdown in the HL-LHC era will alternate between 16 and 20 months. The approximate breakdown of a generic long year is:

- 13 weeks for the Christmas technical stop, including two weeks for hardware commissioning (this would account for three weeks at the end of a year and for 10 weeks at the start of the following year);
- around 160 days proton–proton operation;
- three technical stops of five days duration during the year;
- a four-week ion run (the working assumption is that the ion programme continues into the HL-LHC era; this is not to say that the HL-LHC will operate with ions for its total period of operation);
- time dedicated for special physics and machine development.

A more detailed breakdown is shown in Table 16-1.

Table 16-1: Potential breakdown of a standard HL-LHC year

| Activity | Time assigned [days] |
|---|---|
| Christmas technical stop including hardware commissioning | 91 |
| Commissioning with beam | 21 |
| Machine development | 22 |
| Scrubbing | 7 (to 14) |
| Technical stops | 15 |
| Technical stop recovery | 6 |
| Proton physics including intensity ramp-up | 160 |
| Special physics runs | 8 |
| Ion run setup | 4 |
| Ion physics | 24 |
| Contingency | 7 |

### 16.1.2.2 Availability and operation efficiency

The HL-LHC will place challenging demands on availability and operational efficiency if the ambitious integrated luminosity goals are to be met. Given that the requisite high intensity beam can be injected, ramped, squeezed, collided, and the luminosity levelled, the principle operational challenges will be assuring high availability and maximizing physics efficiency.

**16.1.2.3 Faults**

Faults cover an enormous range from a simple front-end computer reboot, to the loss of a cold compressor with corresponding loss of time, to operations ranging from 10 minutes to potentially days. The impact of a typical fault requiring tunnel access shows the following sequence of steps from the occurrence of the original fault through to full recovery.

- Premature beam dump in stable beams.
- Original diagnosis of fault by control room operator – contact expert.
- Remote diagnosis by expert.
- Access required and prepared.
- Travel of expert to site. Travel of radiation protection piquet to site.
- Intervention, on-site diagnosis, and repair by expert.
- Recovery from access.
- Recovery from the impact of the fault (for example, cool-down following quench).
- Re-establish machine state: pre-cycle, injection, etc.

It can be seen that, besides the cost of fixing a fault, there are also significant additional overheads. Faults often dump the beam with the cost of at least a full turnaround to re-establish the state of the machine. The cost in turnaround time and the fault recovery component are examined in more detail below.

**16.1.2.4 Turnaround time**

A breakdown of the foreseen HL-LHC turnaround time is shown below.

Table 16-2: Breakdown of turnaround and estimated minimum time for each phase

| Phase | Time [minutes] |
|---|---|
| Ramp-down/pre-cycle | 60 |
| Pre-injection set-up | 15 |
| Set-up with beam | 15 |
| Nominal injection | 20 |
| Prepare ramp | 5 |
| Ramp | 25 |
| Flat-top | 5 |
| Squeeze | 25 |
| Adjust/collide | 10 |
| Total | 180 |

From Table 16-2, one can see that realistically a three-hour minimum turnaround time may be assumed in the HL-LHC era. The main components are the ramp-down from top energy, the injection of beam from the SPS, the ramp to high energy, and the squeeze. The ramp-down, ramp, and squeeze duration are given by the current rate limitations of the power converters. Of note is the 10 A/s limit up and down for the main dipoles, the need to respect the natural decay constants of the main quadrupoles, the individually powered quadrupoles, and the triplets during the ramp-down and the squeeze. These quadrupoles are powered by single quadrant power converters, and it takes a considerable time for the current to fall to the required level. A faster pre-cycle via upgrades to the power converters might be anticipated. Two quadrant power converters for the inner triplets, for example, would remove them as a ramp-down bottleneck.

In practice, the turnaround has to contend with a number of issues that could involve lengthy beam-based set-up and optimization. Typical beam-based optimization might include the need to re-steer the transfer lines, occasional energy matching between the SPS and LHC, and the need for the SPS to adjust scraping during the injection process. Injector and LHC tuning and optimization are accounted for in the average turnaround time.

**16.1.2.5 Lost fills and fill length**

Another important consideration directly impacting operational efficiency are the overheads of losing a fill (in the ramp, in the squeeze, in physics). In 2012, for example, 70% of all fills were terminated by a fault with costs ranging from a full turnaround plus fault recovery time to curtailed stable beam time. In 2012, having made it into stable beams, operations experienced a lot of short unproductive fills because of premature dumps. The cost of the short fills is a corresponding number of extra turnarounds that directly lead to lost time for physics. All means must be used during Run 2 and Run 3 to target all causes of premature dumps. In the HL-LHC era a choice of robust operating parameters must be made to avoid losing fills to high beam loss and instabilities during the cycle.

16.1.3 Discussion

A number of studies have been made that model the potential integrated luminosity that can be delivered by the HL-LHC at a levelled luminosity of $5 \times 10^{34}$ cm$^{-2}$ s$^{-1}$. All conclude that it will be challenging to deliver 250 fb$^{-1}$ per year. Every attempt must therefore be made to target high availability, shorter turnaround times, and a low number of premature dumps. Some of the main areas that might be targeted in the interest of improved availability and operational efficiency have been identified.

- Reduce the number of faults (hardware and software) – this is the natural target for improved availability. It requires targeted improvements and consolidation across all systems.
- Reduce time to fix faults, reduce intervention times, and reduce the number of tunnel interventions (for example by universal remote reset functionality, improved remote diagnostics, or increased redundancy).
- Reduce number of beam-induced faults (R2E, beam-induced heating, vacuum issues).
- Reduce the mean turnaround time. Here it is possible to imagine targeting routine optimization, test runs, and the nominal cycle.
- Choose a robust set of beam parameters to avoid unnecessary dumps from beam-related issues.

**16.2 Hardware commissioning**

16.2.1 Commissioning of the superconducting circuits

**16.2.1.1 Testing of the superconducting links in short-circuit configuration**

All superconducting links [1] will be individually (cryogenically and electrically) tested in the SM18 magnet test facility prior to installation in the tunnel [2]. As is the case for all superconducting elements installed in the machine, these links will have to be further tested in situ to assess that no degradation occurred during transport and installation.

A test similar to those currently performed on the warm part of all superconducting circuits is suggested, with one extremity in short-circuit configuration: the electrical connections to the power converters are the nominal ones; the conductor at the other extremity of the link is not connected to the magnets, but short-circuited, in a configuration that allows powering up to the nominal current of all busbar pairs. The link will have to be cooled down to its nominal temperature; thus the configuration and the precise place of short-circuit installation will be decided in accordance with the final design of the connection boxes (DFMs and DFXs), also taking into account the planning constraints.

Once the short-circuit is installed and the nominal cryogenic conditions established, a high voltage qualification (ElQA) of all lines will be performed in order to validate the galvanic insulation versus ground and the capacity of all lines to withstand the mutual high voltages developed during a fast change of current in the different circuits (typically during a fast abort or quench). The values of voltages to be applied and the maximum acceptable leakage current values will be defined at a later stage; Ref. [3] will have to be updated correspondingly.

After the ElQA validation, all lines can be powered. The powering will consist of several phases, each of them to be analyzed and approved by the experts, before progressing to the next phase. At each phase the cryogenics team will have to verify the correct behaviour of the cryogenic system and cooling loop. Each circuit should be powered to increasing current levels (e.g. minimum operational current, 25% of nominal, 50% of nominal, and nominal current). At each current level, the powering, protection, and cryogenic systems will be qualified. The detailed steps and qualification criteria will be identical to those prepared for SM18 and will have to include the simultaneous powering of different circuits, and stress tests such as the fast discharge of a line with all of the others powered at nominal current in order to study the cross-talk between them.

### 16.2.2 Electrical Quality Assurance tests

As stated in Ref. [3], the objective of the ElQA tests is to release each individual superconducting circuit for powering, to gather all the necessary electrical parameters for operation, and to track all the data acquired and to manage the related non-conformities.

**16.2.2.1 ElQA at warm**

At the end of the installation and connection of all magnets, resistance measurements and a high voltage qualification of all circuits will be performed: to check whether the circuit is closed, determine a reference resistance value at warm, and to validate the galvanic insulation versus ground. The values of voltages to be applied and the maximum acceptable leakage current values will be defined at a later stage.

**16.2.2.2 ElQA at cold**

Similar tests will be performed at cold, with larger test voltages applied. The circuits and the corresponding link will be cooled down to their nominal temperature. For the high voltage qualification of all lines, this will be performed to validate the galvanic insulation versus ground and the capacity of all lines to withstand the mutual high voltages developed during a fast change of current in the different circuits (typically during a fast abort or quench).

The high voltage qualification also includes testing of all the elements that are electrically connected to the tested circuit. Such elements are:

- the instrumentation and feedthrough systems;
- the magnet protection units;
- the temperature sensors with the related tunnel cabling and electronics;
- the tunnel cabling for routing the voltage taps used for the protection of the superconducting circuits.

In addition, transfer function measurements will be performed, with the aim of determining the impedance as a function of the frequency. The results of these measurements are used to spot possible inter-turn shorts, and by the power converter group to adjust the power converter regulation.

The values of voltages to be applied and the maximum acceptable leakage current values will be defined in a later stage and Ref. [3] will be updated accordingly.

### 16.2.3 Powering tests

The HL-LHC magnets present several peculiarities [4] that have to be kept in mind for their commissioning. The most relevant are: the fact that all magnets will be cooled down to 1.9 K (instead of 4.5 K as are all present matching section magnets); that $Nb_3Sn$ will be used for the first time; that the current of the inner triplet and the Q4 magnet will be the highest in the machine (close to 20 kA); and, importantly, that some of the high current magnets will be protected only via energy extraction in a dump resistor without quench heaters. In addition, the powering scheme of the inner triplet will be different from the present one, with Q1 in series with Q3; and Q2a and Q2b in series in a separate circuit: this is important in case of a quench of one of the magnets.

The powering of all circuits up to nominal current will be done in steps. At the end of each step, online and offline analyses are performed by equipment owners and protection experts to assess the performance of all hardware in the circuit. In particular, for the powering of individual circuits, several cycles at different current levels will be performed to study the performance of the magnets, the efficiency of the protection mechanisms (by provoking fast aborts and even quenches), and to checks all functionalities of the powering interlocks and of the power converters (via provoked powering failures).

A typical series of tests includes:

- at minimum operational current, testing of the full interlock chain, with the verification of cryogenic signals, power permit, powering failure, circuit quench transmission, and fast power abort requests;

- at low current, a check of the power converter performance and verification of all protection functionalities, by means of provoked slow and fast power aborts, with energy extraction;

- repetition of a series of power aborts and, possibly, simulation of quenches from progressively higher current levels, with more and more energy stored (e.g. 25%, 50%, and 100% of the nominal current).

Before starting a new powering test, all previous tests must have been validated. The validation includes approval by power converter and powering interlock experts, magnet owners, and protection experts. Cryogenics experts should also confirm the correct operation of their installations and instrumentation. The criteria for approval, the parameters, and the relevant information to be stored will be discussed in due time. The first time that these will be applied is in 2019–2020, when the test of a full string of magnets (reproducing from Q1 to D1) is foreseen, powered by a superconducting link. All valuable data extracted from the test on the string will be translated in powering procedure steps and criteria.

After the individual test of all circuits up to the design current, the common powering of a set of circuits will be done for magnets that are in the same cryogenic envelope and are powered from the same link (usually referred to as the powering of a group of circuits). The objective of this simultaneous powering is to validate operation of all magnets in nominal conditions; current cycles similar to those applied in normal operation should be used for the powering of a group of circuits. Important at this stage is the behaviour with combined powering in critical conditions, such as the fast power abort of a circuit when the others are at full current. For the inner triplets, in particular, quenching in a triplet quadrupole might induce a quench in a nearby quadrupole or corrector if the current in this related circuit is not extracted fast enough. These tests should be performed on all the magnets and could well trigger the change of detection thresholds and protection configurations. Once more, all tests should be approved by a group of experts and recorded for future reference.

Particular attention also has to be paid to those circuits that are not equipped with heaters and are protected by energy extraction on a dump resistor. For such circuits, a precise estimate of the energy deposited during a quench has to be made, not only in the case of bench tests, but also in the more severe conditions of combined powering in the tunnel. Eventually, the protection threshold should be adapted to reduce energy deposition and improve magnet safety during powering.

It is important to observe that the first commissioning campaign will already take place in Long Shutdown 2 (LS2 in 2018–2019), when the 11 T $Nb_3Sn$ magnets in the dispersion suppression region around P2 will be installed. The superconducting link at P7 will be installed in the same period to power the circuits presently powered from power converters installed in RR73 and RR77. All powering procedures and commissioning steps will need to be ready by that time.

### 16.3 Hardware commissioning of the HL collimation system

The collimator settings, controls, and operational sequences should be intensively tested during the hardware commissioning phase [5]. A dedicated test to address the reproducibility of collimator movements during critical operational sequences (such as the ramp) will be performed. At this stage the collimators should have been fully installed and the local collimator controls in the tunnel fully validated.

Before beam is injected into the machine, the machine protection (MP) functionality of the collimation system must be guaranteed. Each collimator is connected to the beam interlock system (BIS) and has more than 20 interlocks that will need to be verified. The jaw positions and collimator gaps are monitored via six linear variable differential transformer (LVDT) sensors. These signals are interlocked with inner and outer limit values, making a total of 12 interlocks per collimator. In addition, there are a total of six energy-dependent and $\beta^*$-dependent limit functions and an interlock to protect from 'local' mode collimation control. The temperature of the collimators is also monitored and interlocked with minimum and maximum adjustable thresholds independently for five sensors per collimator. After successful results from these tests, the system will be ready to allow beam into the machine.

The main upgrade of the collimation system for HL-LHC [6] will ensure cleaning of beam halo and will keep losses in high luminosity experimental regions at an appropriate level. For this, the project foresees the installation of local collimation in the dispersion suppressors (DS). The design of a collimator to be installed between two short 11 T dipoles (TCLD) is ongoing. These collimators will feature the latest design improvements, including embedded BPMs for fast alignment. Altogether, between 10 to 14 TCLDs are foreseen for the HL-LHC in addition to eight new tertiary collimators and some additional secondary collimators. Depending on the final update programme, it should be possible to expect about 30 more collimators with respect to the present LHC system. Unlike other hardware commissioning tests (such as the magnets), most of the collimation commissioning can be done parasitically, the main exception being the testing of the interlock system where the BIS needs to be available.

### 16.4 Commissioning of the cryogenic systems

The HL-LHC foresees numerous modifications of the cryogenic system [7]. Among them are:

- the new cooling system for the superconducting links;
- the modifications to cool the elements that previously were at 4.5 K to 1.9 K;
- the independent RF cooling system;
- the cooling loop for the crab cavities, and much more.

The operation of all of them, together with the time needed to qualify and tune the systems, will be detailed once the design is definitive. Provisionally, an approximate time of three weeks is considered to be mandatory to commission the scheme for the superconducting magnets.

### 16.5 Commissioning of the crab cavities

As for all elements in the LHC, the crab cavities will be first tested on the surface to nominal specification prior to installation in the tunnel. In addition, some modules will be installed in the SPS for a complete qualification of the standard two-cavity module with LHC-type beams. This is presently planned for the 2017 run after the extended year-end technical stop.

The test of the cavities can be differentiated into the commissioning of the RF cavities and associated ancillary systems (cryogenics, vacuum), and RF commissioning with beam. The requirements in terms of time and manpower will be assessed later.

Concerning the commissioning of the cryogenics, the correct operation of the cooling loop and the capacity will be verified, together with the expected behaviour of the instrumentation; proper verification criteria and sequence will be defined at later time with input from the qualification of the cryogenic-module on the surface tests. The vacuum integrity and the vacuum interlocks will be tested as well, which should cut the RF power in case of issues and during cavity conditioning.

The conditioning of the cavities will first be performed on the surface in the nominal configuration, but the commissioning of the low-level RF system (the tuning control, the regulation loop around the amplifier, plus the RF feedback) will have to be validated in the SPS for the first time in its nominal configuration. A detailed procedure for the verification of all functionalities will be prepared well in advance. The information and issues arising from these tests will directly help in the efficient commissioning of the system in the LHC.

All possible RF manipulations foreseen for the LHC operation cycle will first be performed without beam. An important verification concerns the efficiency of the fast feedback of the cavity field. The delay loops in the SPS between the two cavities will be arranged to mimic the cavity setup in the LHC to both ensure the fast and independent control of the cavity set point voltage and phase, and the slower loop to regulate the cavities on either side of the IP. This is essential to ensure machine protection in the event of an abrupt failure of one of the cavities.

**16.6 Commissioning with beam**

One is fully able to draw on past experience in outlining a commissioning plan for initial operation following LS3. A skeleton plan is shown in Table 16-3.

Table 16-3: Outline of initial commissioning following LS3

| Phase | Key objectives |
|---|---|
| Injection and first turn | Injection region aperture, injection kicker timing |
| Circulating beam | RF capture, beam instrumentation, initial parameter checks |
| 450 GeV initial commissioning | Transfer line and injection set-up, orbit, |
| 450 GeV measurements and setup | Beam instrumentation, optics, aperture, collimation, LBDS |
| 450 GeV two-beam operation | Separation bump checks, beam instrumentation |
| Ramp | Snapback, chromaticity control, orbit and tune feedbacks |
| Flat-top checks | Collimation, optics, orbit, decay |
| Squeeze | Optics, collimation set-up |
| Validation | Loss maps, asynchronous dumps |
| Collide | First stable beam with a low number of bunches |

The initial commissioning phase should evolve through initial set-up, system commissioning through the nominal cycle, standard measurement and correction, set-up of protection devices, and validation. It is a relatively complex phase with necessary interplay between the various teams to allow beam-based commissioning of systems such as tune and orbit feedbacks, transverse dampers, RF, etc. under appropriate conditions at the various phases of the operational cycle.

The aims of the initial commissioning phase are as follows.

- Establish nominal cycle with a robust set of operating parameters. This will include commissioning of the squeeze to an appropriate β* with measurement and correction of the optics and key beam parameters at each stage. One should not expect to probe the limits of the HL-LHC parameter space at this stage.

- Measure and correct the optics. Measure the aperture.

- Set-up injection, beam dump, and collimation, and validate set-up with beam.

- Commission beam-based systems: transverse feedback, RF, injection, beam dump systems, beam instrumentation, and orbit and tune feedbacks.

- Commission and test machine protection backbone with beam.

- Check the understanding of: magnet model; higher order optics.

The initial commissioning phase is performed at low intensity, with a low number of bunches, and a generally safe beam. The output of this phase is taken to be first collisions in stable beams with a small number of bunches. Following this, pilot physics can be delivered with up to 100 widely space bunches. Scrubbing will then be required before entering the intensity ramp-up phase. Scrubbing could well follow the two-stage approach deployed following LS1. This approach is outlined below.

- Initial scrubbing with 50 ns and 25 ns beam following initial commissioning opening the way for a period of 50 ns operation.

- Intensity ramp-up with 50 ns is then foreseen. During this stage system commissioning with higher intensity continues (instrumentation, RF, TFB, injection, beam dumps, machine protection, vacuum, etc.). Variables at this stage include: bunch intensity, batch structure, number of bunches, and emittance. Physics fills can be kept reasonably short. The intensity ramp-up is performed in a number of clearly defined steps with associated machine protection and other checks.

- Following the 50 ns intensity ramp-up a reasonably short period of 50 ns operation is envisaged. This will be used to characterize vacuum, heat load, electron cloud, losses, instabilities, UFOs, and impedance.

- Thereafter a further scrubbing period with 25 ns and the doublet beam will be required for 25 ns operation.

- This is followed by an intensity ramp-up with 25 ns dictated by electron cloud conditions, with further scrubbing as required.

Important beam-related characteristics such as lifetime, beam loss through the cycle, stability, quench levels, and UFO rates will only become accessible with an increase in bunch intensity and number of bunches during the intensity ramp-up phase.

### 16.6.1 Operation with heavy ions

In a typical LHC operating year, one month is reserved for the nuclear collision programme.

#### 16.6.1.1 Pb–Pb operation

Collisions of fully stripped $^{208}$Pb$^{82+}$ nuclei are requested by the ATLAS, ALICE, and CMS experiments. These heavy-ion runs will generally be scheduled following an extended period of p–p operation. The reduced activation of the LHC and injectors will allow interventions to start promptly in the subsequent shutdown periods. The runs will be short, typically of a total of one month, and there is a high premium on rapid commissioning of collision conditions with full luminosity within a few days.

To minimize commissioning time, the heavy-ion magnetic cycle will generally exploit the magnetic reproducibility of the LHC and equal magnetic rigidity of the beams to use as much as possible of the cycle previously established for proton beams. Due care must nevertheless be taken to reproduce the same orbits in the higher-sensitivity mode of the beam position monitors and to adjust the RF frequencies to capture the ion beams [7]. However the collision optics will always be different and a new squeeze process will have to be established to provide a low $\beta^* = 0.5$–$0.6$ m at the ALICE experiment with similar values (generally not as low as for protons) at ATLAS and CMS. Crossing angles will also be changed, with that of ALICE carefully minimized for the benefit of its zero-degree calorimeter. With the 50 ns bunch spacing, the minimum acceptable separations at parasitic encounters will be established empirically [9, 10]. In all cases, the validation of the collimation configuration and machine protection will be an essential step in the set-up of heavy-ion physics conditions. During each run, a reversal of the ALICE spectrometer magnet polarity and flipping of the crossing angle should be foreseen [9].

Controlled blow-up of the longitudinal emittance, using RF noise with a spectrum focussed on the core of the bunch, will play an important role in minimising transverse emittance blow-up due to intra-beam scattering at injection [10] and, very likely, also during stable beams.

Because of the very rapid luminosity burn-off (halving in about two hours with three experiments active), the fills will be short and frequent refilling and recycling of the machine will be required, typically four times per day. Although the experiments will be able to handle the highest peak luminosity foreseen, it may nevertheless be useful to briefly level the luminosity with separation at the start of fills to reduce the intensity burn-off during the initial set-up of collision conditions. This will have negligible impact on average daily luminosity.

Secondary beams created in the collisions [10, 11] will impinge on the beam pipe in the dispersion suppressors on either side of each collision point. The power in these beams is expected to be sufficient to quench a superconducting magnet, and counter-measures will have to be applied immediately when the beams are put into collision. The most effective are expected to be the dispersion suppressor collimators [10] which will be positioned during the squeeze process. Initially these will be installed in IR2 for the ALICE experiment. The attainable peak luminosity for ATLAS and CMS will be maximized using an orbit bump technique that will spread the losses over a larger area. If necessary, similar collimators may be installed in IR1 and IR5 later. In any case, the dump thresholds for the beam-loss monitors in the dispersion suppressors will have to be carefully adjusted.

Because of the more complicated nuclear and electromagnetic processes that can occur when lead ions interact with the material of the collimators, collimation efficiency will be lower than for protons and quite different loss patterns are expected [12]. Here, again, the energy deposition from the losses may approach or exceed quench levels and may necessitate the installation of dispersion suppressor collimators in IR7.

### 16.6.1.2 p–Pb operation

Asymmetric collisions of protons with $^{208}$Pb$^{82+}$ nuclei are requested by the ATLAS, ALICE, CMS, and LHCb experiments. The smaller forward detectors installed around the ATLAS and CMS interaction points also study these collisions. These runs are scheduled in place of the Pb–Pb runs, to a similar timetable, during a fraction of the operating years.

The feasibility of this complex mode of operation, unforeseen in the original LHC baseline design, was demonstrated by the first runs in 2012 and 2013 [13]. Proton beams will be injected first since the Pb beams suffer more from intra-beam scattering at injection energy. The RF frequencies corresponding to capture of the two beams on their central orbits are significantly different at injection energy, requiring the RF systems of the two rings to operate independently. At collision energy the frequencies will be locked together, displacing the proton orbit outwards and the Pb orbit inwards, by a small fraction of a millimetre, onto off-momentum orbits. A 'cogging' procedure, again using RF frequencies, will be used to move the collision points to their proper locations in the experiments [14].

Proton–lead runs place a number of requirements on the filling schemes. The basic bunch spacing must be created by both the heavy-ion and proton pre-injector chains to ensure that the bunch patterns largely match each other. The trains must be distributed to provide an appropriate share of collisions in each experiment. As seen in 2013, this can result in a complicated set of beam–beam encounter classes, bunch parameters, and lifetimes. In general, the proton beam size at the collision points will be smaller than the Pb beam, a circumstance that further reduces the lifetime of the Pb beam. Mitigation of this problem would be a second, less obvious, benefit of a stochastic cooling system.

As in Pb–Pb operation a new squeeze process will have to be commissioned, including all four experiments, with different end-points. This will generally be prepared and corrected first, using on-momentum proton beams. Pre-computed chromatic corrections [14] will be incorporated for the final off-momentum squeeze.

Rapid luminosity burn-off of the Pb beam on the proton beam will determine the length of fills. The luminosity will be essentially proportional to the Pb beam intensity, and the integrated luminosity delivered per fill summed over all experiments will be a fraction of $N_{Pb}/\sigma_b$ where $N_{Pb}$ is the total number of Pb ions in the beam at the start of collisions and $\sigma_b$ is the cross-section for luminosity burn-off in p–Pb collisions. The role of the proton bunch intensity is only to determine the lifetime of the Pb beam and the length of the fill. Values a few times less than in proton operation produce fills of reasonable length and satisfy operational constraints related to the functioning of the beam position monitors with unequal beams.

### 16.6.1.3 Other species

Collisions of lighter nuclei may eventually be requested by the experiments. At the time of writing, the injector chain is being commissioned with argon ions (for fixed target operation) for the first time. When this process is well-advanced, the beam quality and intensity data will allow performance projections for Ar–Ar and p–Ar collisions in the LHC.

Since most of the performance limitations for ion beams depend on high powers of the nuclear charge as well as beam intensity, their relative importance may shift considerably. First indications are that collimation losses from lighter ions will be more of a concern than for lead, while losses from collision processes will be less serious.